\documentclass[aps,prl,twocolumn,superscriptaddress,showpacs,preprintnumbers,draft]{revtex4}

\usepackage{array}
\usepackage{dcolumn}

\newcommand{\slj}[3]{\mbox{$^{#1}${\ifcase#2\or S\or 
         P\or D\or F\fi}$_{#3}$}}
\newcommand{\sLj}[3]{{}^{#1}\!#2_{#3}}
	 
\newcommand{\kev}{\hbox{ keV}}
\newcommand{\mev}{\hbox{ MeV}}
\newcommand{\gev}{\hbox{ GeV}}

\newcommand{\cfrac}[2]{\textstyle{\frac{#1}{#2}}}

\newcommand{\jpsi}{\ensuremath{J\!/\!\psi}}

\begin{document}

\preprint{FERMILAB--PUB--02/104--T}
\preprint{BUHEP-02-25}
%\preprint{hep-ph/0206018}

\title{$B$-Meson Gateways to Missing Charmonium Levels}


\author{Estia J. Eichten}
\email[E-mail: ]{eichten@fnal.gov}
%\thanks{}
\affiliation{Theoretical Physics Department\\ Fermi National 
Accelerator Laboratory\\ P.O.\ Box 500, Batavia, IL 60510}
\author{Kenneth Lane}
\email[E-mail: ]{lane@physics.bu.edu}
%\thanks{}
\affiliation{Theoretical Physics Department\\ Fermi National 
Accelerator Laboratory\\ P.O.\ Box 500, Batavia, IL 60510}
\affiliation{Department of Physics, Boston
University\\ 590 Commonwealth Avenue, Boston, MA 02215}
\author{Chris Quigg}
\email[E-mail: ]{quigg@fnal.gov}
%\thanks{}
\affiliation{Theoretical Physics Department\\ Fermi National 
Accelerator Laboratory\\ P.O.\ Box 500, Batavia, IL 60510}


\date{\today}

\begin{abstract}
We outline a coherent strategy for exploring the four remaining narrow
charmonium states [$\eta_{c}^{\prime}(2\slj{1}{1}{0})$,
$h_{c}(1\slj{1}{2}{1})$, $\eta_{c2}(1\slj{1}{3}{2})$, and
$\psi_{2}(1\slj{3}{3}{2})$] expected to lie below charm threshold. 
Produced in $B$-meson decays, these levels should be identifiable
\textit{now} via striking radiative transitions among charmonium levels
and in exclusive final states of kaons and pions. Their
production and decay rates will provide much needed new tests
for theoretical descriptions of heavy quarkonia.
\end{abstract}

\pacs{14.40.Nd,14.40.Gx,13.25.Gv}
%\keywords{}

\maketitle

%\section{}
%\subsection{}
%\subsubsection{}
Every new spectroscopy carries the potential to illuminate older
spectroscopies in novel---and sometimes unexpected---ways.  The decays
of charmed mesons have offered new paths to the study of
mesons---especially excited states---composed of light quarks, and
decays of charmonium ($c\bar{c}$) states have provided new access to
glueballs---hadrons composed largely of gluons.  The study of
semileptonic $\tau$ decays has
considerably enriched our knowledge of $a_{1}$ properties.  $B$-meson
decays into charmonium states are an indispensable tool for the
exploration of \textsf{CP}
violation~\cite{Harrison:1998yr,Anikeev:2001rk}.  They can also serve
as gateways to the discovery of hitherto unobserved charmonium states. 
The properties of these missing states can illuminate the interquark
force and reveal effects that lie outside the simple quarkonium
potential framework, including the influence of virtual decay
channels. 

Detailed knowledge of the $c\bar{c}$ spectrum first derived from the
study of $e^{+}e^{-}$ annihilations, which explored the \slj{3}{1}{1} 
and \slj{3}{3}{1}
$J^{PC} = 1^{--}$ states and the \slj{3}{2}{J} or \slj{1}{1}{0} states
connected to them by E1 or M1 radiative transitions. Though incisive, 
the $e^{+}e^{-}$ annihilation channel has limitations. Twenty years 
have passed without a confirmation of the Crystal Ball claim of the 
2\slj{1}{1}{0} $\eta_{c}^{\prime}(3594\pm5)$ 
\cite{Edwards:1982mq}, and the complementary technique of charmonium 
formation in $p\bar{p}$ annihilations does not support the 
$\eta_{c}^{\prime}(3594)$ observation \cite{Ambrogiani:2001wg}. 
The sighting of the 1\slj{1}{2}{1} level near the \slj{3}{2}{J} center 
of gravity in $p\bar{p} \rightarrow 
h_{c}(3526) \rightarrow \pi^{0}\jpsi$ reported 
by Fermilab Experiment E760~\cite{Armstrong:1992ae} needs 
confirmation.  The
$\eta_{c2}(1\slj{1}{3}{2})$ and 
$\psi_{2}(1\slj{3}{3}{2})$ have also proved elusive, with only an
unsubstantiated claim in the literature for a $2^{--}$ state,
$\psi(3836\pm13)$, in $\pi^{\pm}N \rightarrow \jpsi\pi^{+}\pi^{-} + 
\hbox{anything}$ \cite{Antoniazzi:1994jz}.

Help is on the way.  The CDF experiment's observation of $B$ decays to
known charmonium levels shed light on prompt and secondary production
of ($c\bar{c}$) states, sharpening the puzzle of the production
mechanism \cite{Papadimitriou:2001bb}.  The $e^{+}e^{-} \rightarrow
\Upsilon(\mathrm{4S}) \rightarrow B\bar{B}$ experiments
CLEO~\cite{Edwards:2000bb}, BaBar~\cite{Aubert:2001xs}, and
Belle~\cite{Abe:2001pf}, report increasingly detailed
observations of the established quarkonium states in $B$ decay.  And now
Belle has reported the observation of the 2\slj{1}{1}{0}
($\eta_{c}^{\prime}$) level at a new mass (3654\mev) in exclusive $B
\rightarrow KK_sK^-\pi^+$ decays \cite{newbelle}.

In this paper, we give a template for the spectrum of charmonium based
on a Coulomb + linear potential, and we present estimates of the
principal decay rates for the unobserved states to make quantitative
the expectation that four narrow states remain to be studied.  We
argue that we may expect ample production of the missing charmonium
states in $B$-meson decays, and we suggest experimental strategies for
detecting the missing levels.  As with the discoveries of the
P--states $\chi_{cJ}$ and $\eta_{c}$ of charmonium, radiative
transitions will be of central importance.  Finally, we comment on
what we will learn by studying the masses and properties of the
missing levels.

The importance of radiative decays to the discovery of charmonium 
levels, including the D-wave states, has been appreciated since the 
earliest days of charmonium spectroscopy~\cite{Eichten:1975af}. 
Recently, Ko, Lee and Song~\cite{Ko:1997rn} discussed
the observation of the narrow D states by photonic and pionic
transitions, and Suzuki~\cite{Suzuki:2002sq} emphasized that the cascade
decay $B \rightarrow h_c K^{(*)} \rightarrow \gamma \eta_c K^{(*)}$
offers a promising technique to look for $h_{c}$.

\textit{The spectrum of charmonium.} To estimate the positions of the
missing charmonium levels, we have adjusted parameters of the classic
Cornell potential \cite{Eichten:1978tg,Eichten:1980ms},
\begin{equation}
    \begin{array}{c}
	V(r)=-\displaystyle{\frac{\kappa}{r} + \frac{r}{a^2}}\;\;,\\[9pt]
	m_c = 1.84\gev,\;\;\kappa=0.61,\;\;a=2.38\gev^{-1},
    \end{array}
	\label{eq:cornellpot}
\end{equation}
to reproduce the observed centers of gravity of the 1S and 1P states
\footnote{For quarks bound in a central potential, it is convenient to
separate the Schr\"{o}dinger wave function into radial and angular
pieces as $\Psi_{n\ell m}(\vec{r}) = R_{n\ell}(r)Y_{\ell
m}(\theta,\varphi)$, where $n$ is the principal quantum number, $\ell$
and $m$ are the orbital angular momentum and its projection,
$R_{n\ell}(r)$ is the radial wave function, and $Y_{\ell
m}(\theta,\varphi)$ is a spherical harmonic.  It is also useful to
introduce the reduced radial wave function, $u_{n\ell}(r) =
rR_{n\ell}(r)$.}.  No one has produced a satisfactory analytic (or
potential-based) explanation of the spin splittings of the
1\slj{3}{2}{J} levels.  Moreover, the 2S, 1D, and 2P levels are
certain to be influenced appreciably by coupling to decay channels. 
Accordingly, we will not estimate the spin splittings of those levels
beyond offering the expectation that they will be small.  Our expectations for the charmonium spectrum
are summarized in Table~\ref{table:twlevels}.

The 1\slj{1}{2}{1} $h_{c}$ and the 2\slj{1}{1}{0} $\eta_{c}^{\prime}$
of course lie below $D\bar{D}$ threshold, and so will be typically
narrow charmonium states.  In the absence of strong influence from the
coupling to decay channels, the 2\slj{3}{2}{J} $\chi_{c}^{\prime}$ and
2\slj{1}{2}{1} $h_{c}^{\prime}$ states should lie well above the
$D\bar{D}$ and $D^{*}\bar{D}$ thresholds, and so should have
uninhibited strong decays.  As has long been known, the $J^{PC} =
2^{-+}$ 1\slj{1}{3}{2} $\eta_{c2}$ and $J^{PC} = 2^{--}$
1\slj{3}{3}{2} $\psi_{2}$ states constitute an important special
case: they lie between the $D\bar{D}$ and $D^{*}\bar{D}$ thresholds,
but are forbidden (because of their unnatural parity) to decay into
$D\bar{D}$.  It is therefore plausible that they will appear as narrow
levels, and we now quantify this suspicion.

\newcolumntype{+}{D{:}{ \pm }{-4}}

\begin{table}%[H] add [H] placement to break table across pages
\caption{$c\bar{c}$ spectrum in the Coulomb $+$ linear  potential 
{\protect (\ref{eq:cornellpot}}).\label{table:twlevels}}
\begin{ruledtabular}
    \begin{tabular}{c l d +}
	\multicolumn{2}{c}{State\phantom{MMMM}} & \multicolumn{1}{c}{\text{Mass (MeV)}}   & 
	\multicolumn{1}{c}{\text{Remarks}} \\
	\hline
$\begin{array}{c}
    1\slj{1}{1}{0} \\
    1\slj{3}{1}{1} 
\end{array}$ &
$\left.\begin{array}{c}
    \eta_{c} \\
    \jpsi 
\end{array}\right\}$ c.o.g. & 3067.\footnotemark[1] & 
\multicolumn{1}{c}{$\left\{
\begin{array}{c}
    2979.8 \pm 1.8 \\
    3096.87 \pm 0.04
\end{array}
\right.$} \\
1\slj{1}{2}{1} & $h_{c}$ & 3526.   & \\
$\begin{array}{c}
    1\slj{3}{2}{0}  \\
    1\slj{3}{2}{1}  \\
    1\slj{3}{2}{2}
\end{array}$ &
$\left.\begin{array}{c}
    \chi_{c0}  \\
    \chi_{c1}  \\
    \chi_{c2}
\end{array}\right\}$ c.o.g. & 3526.\footnotemark[1] & 
\multicolumn{1}{c}{$\left\{
\begin{array}{c}
    3415.0 \pm 0.8  \\
    3510.51 \pm 0.12  \\
    3556.18 \pm 0.13
\end{array}
\right.$} \\
$\begin{array}{c}
    2\slj{1}{1}{0} \\
    2\slj{3}{1}{1} 
\end{array}$ &
$\left.\begin{array}{c}
    \eta_{c}^{\prime} \\
    \psi^{\prime} 
\end{array}\right\}$ c.o.g. & 3678. & 
\multicolumn{1}{c}{$\left\{
\begin{array}{c}
    3654. \pm 10. \\
    3685.96 \pm 0.09
\end{array}
\right.$} \\
1\slj{1}{3}{2} & $\eta_{c2}$ & 3815.   & \multicolumn{1}{c}{\text{$\not\rightarrow D\bar{D}$ 
(parity)}}\\
$\begin{array}{c}
    1\slj{3}{3}{1}  \\
    1\slj{3}{3}{2}  \\
    1\slj{3}{3}{3}
\end{array}$ &
$\left.\begin{array}{c}
    \psi  \\
    \psi_{2}  \\
    \psi_{3}
\end{array}\right\}$ c.o.g. & 3815. & 
\multicolumn{1}{c}{$\left\{
\begin{array}{c}
    3769.9 \pm 2.5  \\
    \text{$\not\rightarrow D\bar{D}$ (parity)} \\
    \text{$\rightarrow D\bar{D}$} 
\end{array}
\right.$} \\
% & & & \\
2P &  & 3968.  & \text{\phantom{M}}  \\
1F &  & 4054.  & \text{\phantom{M}}  \\
3S &  & 4118.  & \text{\phantom{M}}  \\
\hline
% & & & \\
\multicolumn{2}{c}{$D^{0}\bar{D}^{0}$} & 3729.0 & \multicolumn{1}{c}{\text{threshold}} \\
\multicolumn{2}{c}{$D^{+}D^{-}$} & 3738.6 & \multicolumn{1}{c}{\text{threshold}} \\
\multicolumn{2}{c}{$D^{0}\bar{D}^{*0} \hbox{ or } D^{*0}\bar{D}^{0}$} & 3871.2 & \multicolumn{1}{c}{\text{threshold}} \\
\multicolumn{2}{c}{$D^{\pm}D^{*\mp}$} & 3879.3 & \multicolumn{1}{c}{\text{threshold}} \\
\multicolumn{2}{c}{$D_{s}^{+}D_{s}^{-}$} & 3973.2 & \multicolumn{1}{c}{\text{threshold}} \\
\multicolumn{2}{c}{$D^{*0}\bar{D}^{*0}$} & 4013.4 & \multicolumn{1}{c}{\text{threshold}} \\
\multicolumn{2}{c}{$D^{*+}D^{*-}$} & 4020.0 & \multicolumn{1}{c}{\text{threshold}} \\
\multicolumn{2}{c}{$D_{s}^{+}\bar{D}_{s}^{*-} \hbox{ or } 
D_{s}^{*+}\bar{D}_{s}^{-}$} & 4099.0 & \multicolumn{1}{c}{\text{threshold}} \\
\multicolumn{2}{c}{$D_{s}^{*+}D_{s}^{*-}$} & 4224.8 & \multicolumn{1}{c}{\text{threshold}} \\
    \end{tabular}
\end{ruledtabular}
\footnotetext[1]{Input values.}
\end{table}

\textit{Properties of the missing levels.} To estimate the decay
rates, we shall use the established values for the $\eta_{c}, \jpsi,
\chi_{c}, \psi^{\prime}$, and $\psi(3770)$ states, adopt the
Belle value for $M_{\eta_{c}^{\prime}}$, set $M_{h_{c}} =
3526\mev$, and choose $M_{\eta_{c2}} = M_{\psi_{2}} = 3815\mev$. We 
estimate the rates for hadronic and radiative decays in turn.

Among hadronic decays, we consider transitions ($\pi\pi$ emission) and
annihilations.  To estimate the $\pi^{+}\pi^{-}+\pi^{0}\pi^{0}$
transition rates, we use the standard multipole expansion of the color
gauge field~\cite{Gottfried:1978gp,Voloshin:1979hc,Yan:1980uh} to express the E1-E1 transition
rates through the Wigner-Eckart theorem given in Eqn.~(3.5) of
Ref.~\cite{Eichten:1994gt}, with experimental inputs given in Table X
of that paper.  The results are shown in Table~\ref{table:hadronic}. 
For present purposes, the essential lesson is that we do not expect
the $\pi\pi$ transition rates to be large for the missing levels of
charmonium.

\newcolumntype{w}{D{:}{}{-6}}
\begin{table}%[H] add [H] placement to break table across pages
\caption{Hadronic decay widths of charmonium states.\label{table:hadronic}}
\begin{ruledtabular}
\begin{tabular}{ccw}
    $c\bar{c}$ state & Decay & \multicolumn{1}{c}{Partial Width} \\
    \hline
    1\slj{1}{1}{0} & $\eta_{c} \rightarrow gg$ & 17.4 \pm 2.8:\mev\hbox{ 
    \protect{\cite{Tomaradze:2002}}} \\
    1\slj{3}{1}{1} & $\jpsi \rightarrow ggg$ & 52.8 \pm 5:\kev\hbox{ 
    \protect{\cite{Groom:2000in}}} \\
    1\slj{1}{2}{1} & $h_{c} \rightarrow ggg$ & 720 \pm 
    320:\kev\footnotemark[1] \\
    1\slj{3}{2}{0} & $\chi_{c0} \rightarrow gg$ & 14.3 \pm 3.6:\mev 
    \footnotemark[2] \\
    1\slj{3}{2}{1} & $\chi_{c1} \rightarrow ggg$ & 0.64 \pm 
    0.10:\mev\footnotemark[2] \\
    1\slj{3}{2}{2} & $\chi_{c2} \rightarrow gg$ & 1.71 \pm 
    0.21:\mev\footnotemark[2] \\
    2\slj{1}{1}{0} & $\eta_{c}^{\prime} \rightarrow gg$ & 
    8.3 \pm 1.3:\mev\footnotemark[3]\\
     & $\eta_{c}^{\prime} \rightarrow \pi\pi\eta_{c}$ & 
     160:\kev\footnotemark[4] \\
    2\slj{3}{1}{1} & $\psi^{\prime} \rightarrow ggg$ & 23 \pm 
    2.6:\kev\hbox{ \protect{\cite{Groom:2000in}}} \\
     & $\psi^{\prime} \rightarrow \pi\pi\jpsi$ & 152 \pm 17:\kev\hbox{ 
     \protect{\cite{Groom:2000in}}} \\
     & $\psi^{\prime} \rightarrow \eta\jpsi$ & 6.1 \pm 1.1:\kev\hbox{ 
     \protect{\cite{Groom:2000in}}} \\
    1\slj{1}{3}{2} & $\eta_{c2} \rightarrow gg$ & 
    110:\kev\footnotemark[5] \\
     & $\eta_{c2} \rightarrow \pi\pi\eta_{c}$ & \approx 45:\kev\footnotemark[4] \\
    1\slj{3}{3}{1} & $\psi \rightarrow ggg$ & 216:\kev\footnotemark[6] \\
     & $\psi \rightarrow \pi\pi\jpsi$ & 43 \pm 15:\kev\footnotemark[7] 
     \\
    1\slj{3}{3}{2} & $\psi_{2} \rightarrow ggg$ & 
    36:\kev\footnotemark[6] \\
     & $\psi_{2} \rightarrow \pi\pi\jpsi$ & \approx 
     45:\kev\footnotemark[4] \\
    1\slj{3}{3}{3} & $\psi_{3} \rightarrow ggg$ &102:\kev\footnotemark[6] \\
     & $\psi_{3} \rightarrow \pi\pi\jpsi$ & \approx 
     45:\kev\footnotemark[4] \\
\end{tabular}
\end{ruledtabular}
\footnotetext[1]{Computed from \slj{3}{2}{J} rates using formalism of 
{\protect \cite{Maltoni:2000km}}; also see~\cite{Brambilla:2001xy}.}
\footnotetext[2]{Compilation of data analyzed by Maltoni, Ref.\
{\protect \cite{Maltoni:2000km}}.} %
\footnotetext[3]{Scaled from $\Gamma(\eta_{c} \rightarrow gg)$.}
\footnotetext[4]{Computed using Eqn.~(3.5) of Ref.\ {\protect \cite{Eichten:1994gt}}.}
\footnotetext[5]{Computed using Eqn.~(\ref{eq:1d2ann}).}
\footnotetext[6]{Computed using Eqn.~(\ref{eq:3djann}).}
\footnotetext[7]{From rates compiled in Table X of Ref.\ {\protect 
\cite{Eichten:1994gt}}.} %
\end{table}

For the annihilations into two or three gluons, we use the standard
(lowest-order) perturbative QCD formulas~\cite{Kwong:1988ak} to scale from
available measurements for related states.  This is a straightforward
exercise for the S-wave levels.  We use Maltoni's
analysis~\cite{Maltoni:2000km} of the \slj{3}{2}{J} annihilation rates
to estimate the rate for $h_{c} \rightarrow ggg$.  The rates for
annihilations of the \slj{3}{3}{J} states into three gluons (via
color-singlet operators) are given by
\cite{Belanger:1987cg,Kwong:1988ae}
\begin{equation}
    \Gamma(\slj{3}{3}{J} \rightarrow ggg) =
    \frac{10\alpha_{s}^{3}}{9\pi}\: \mathcal{C}_{J}\:
    \frac{|R_{n2}^{(2)}(0)|^{2}}{m_{c}^{6}} \: 
    \ln4m_{c}\langle{r}\rangle\; ,
    \label{eq:3djann}
\end{equation}
where $R_{n\ell}^{(\ell)} \equiv 
    \left.d^{\ell}R_{n\ell}(r)/dr^{\ell}\right|_{r=0}$, 
$\langle{r}\rangle = \int_0^\infty \!\!\!\!dr\, r \, u_{n\ell}^{2}(r)$,    
and
$\mathcal{C}_{J} = \cfrac{76}{9}, 1, 4$ for $J=3,2,1$.  A
complete analysis (including color-octet operators as well) has too
many unknowns to be of use \footnote{A theoretical (lattice)
calculation of the color-octet matrix elements is needed.}.  The
strengths of the $J=3,2,1$ annihilations are more generally proportional to
$\mathcal{C}_{J}$, even if color-octet operators
dominate~\cite{Huang:1997sw}.
The two-gluon annihilation rate of the \slj{1}{3}{2} state is 
given by~\cite{Novikov:1978dq}
\begin{equation}
     \Gamma(\slj{1}{3}{2} \rightarrow gg) =
    \frac{2\alpha_{s}^{2}}{3}\: 
    \frac{|R_{n2}^{(2)}(0)|^{2}}{m_{c}^{6}} \; .
    \label{eq:1d2ann}
\end{equation}
Our estimates for the annihilation rates are collected in 
Table~\ref{table:hadronic}. The expectation for 
$\Gamma(\eta_{c}^{\prime} \rightarrow gg)$ is to be compared with the 
Belle value of $15 \pm 24\hbox{ (stat) }\mev$~\cite{newbelle}.

The most prominent radiative decays of charmonium states are the E1
transitions, for which the
rate~\cite{Eichten:1977jk,Novikov:1978dq} is given by
\begin{equation}
    \begin{array}{l}
        \Gamma(\hat{n}\:\sLj{2s+1}{\ell}{J}
        \stackrel{\mathrm{E1}}{\longrightarrow}
        \hat{n}^{\prime}\:\sLj{2s+1}{\ell^{\prime}}{J^{\prime}}\,\gamma) =
	\displaystyle{\frac{4\alpha e_{c}^{2}}{3}}(2J^{\prime}+1)k^{3} 
%	\text{\phantom{}}  
        \\[6pt]
        \quad \times|\mathcal{E}_{\hat{n}\ell : \hat{n}^{\prime}\ell^{\prime}}|^{2} \cdot
         \max{(\ell,\ell^{\prime})}
        \left\{
        \begin{array}{ccc}
        J & 1 & J^{\prime}  \\
        \ell^{\prime} & s & \ell
        \end{array}\right\}^{2}\; ,
    \end{array}
    \label{eq:meister}
\end{equation}
where $e_{c} = \cfrac{2}{3}$ is the charm-quark charge, $k$ is the 
photon energy, the E1 transition matrix
element is $\mathcal{E}_{\hat{n}\ell : \hat{n}^{\prime}\ell^{\prime}} =
\cfrac{3}{k}\int_0^\infty \!\!\!\!dr \, u_{n\ell}(r)
u_{n^{\prime}\ell^{\prime}}(r) \left[\cfrac{kr}{2} j_0(\cfrac{kr}{2}) -
j_1(\cfrac{kr}{2})\right] + O(k/m_c)$, $\hat{n} \equiv n - \ell$ is 
the radial quantum number, and $\left\{ : : : \right\}$ is a 6-$j$
symbol. For M1 transitions, the rate is given by
\begin{equation}
    \Gamma(\hat{n}\:\sLj{2s+1}{\ell}{J}
        \stackrel{\mathrm{M1}}{\longrightarrow}
        \hat{n}^{\prime}\:\sLj{2s^{\prime}+1}{\ell}{J^{\prime}}\,\gamma) =
	\frac{4\alpha e_{c}^{2}}{3m_{c}^{2}}(2J^{\prime}+1) k^{3}
	|\mathcal{M}_{\hat{n}\ell : \hat{n}^{\prime}\ell}|^{2} \,,
	\label{eq:m1form}
\end{equation}
where $\mathcal{M}_{n\ell : n^{\prime}\ell} = \int_0^\infty
\!\!\!\!dr \, u_{n\ell}(r) u_{n^{\prime}\ell}(r)
j_{0}(\cfrac{kr}{2})$.

The calculated rates for the prominent transitions among charmonium 
states are shown in Table~\ref{table:radtranstw}. Values enclosed in 
parentheses have been corrected for the effects of coupling to decay 
channels, following the procedure developed in~\cite{Eichten:1980ms}. 
The calculated values reproduce the patterns exhibited by 
measurements, and are in good agreement with other calculations in 
the literature~\cite{Sebastian:1997cy}. We expect them to provide 
reasonable guidance to the radiative decay rates of the missing 
charmonium levels.

\begin{table}%[H] add [H] placement to break table across pages
\caption{Calculated and observed rates for radiative transitions 
among charmonium levels in the potential (\ref{eq:cornellpot}).\label{table:radtranstw}}
\begin{ruledtabular}
\begin{tabular}{lcc+} 
     & $\gamma$ energy& \multicolumn{2}{c}{Partial width (keV)} \\
Transition & $k$ (MeV) & Computed & 
\multicolumn{1}{c}{Measured\footnotemark[1]} \\
\hline
$\psi \stackrel{\mathrm{M1}}{\longrightarrow} \eta_{c}\gamma$ &
115 & 1.92 &  1.13 : 0.41 \\[3pt]
$\chi_{c0} \stackrel{\mathrm{E1}}{\longrightarrow} \jpsi\gamma$ & 
303 & 120~(105)\footnotemark[2]  & 98 : 43 \\
$\chi_{c1} \stackrel{\mathrm{E1}}{\longrightarrow} \jpsi\gamma$ & 
390 & 242~(215)\footnotemark[2]  & 240 : 51 \\
$\chi_{c2} \stackrel{\mathrm{E1}}{\longrightarrow} \jpsi\gamma$ & 
429 & 315~(289)\footnotemark[2]  & 270 : 46 \\
$h_{c} \stackrel{\mathrm{E1}}{\longrightarrow} \eta_{c}\gamma$ & 
504 & 482 & \\[3pt]
$\eta_{c}^{\prime}\stackrel{\mathrm{E1}}{\longrightarrow} 
h_{c}\gamma$ & 126 & 51 & \\[3pt]
$\psi^{\prime} \stackrel{\mathrm{E1}}{\longrightarrow} 
\chi_{c2}\gamma$ & 128 & 29~(25)\footnotemark[2] & 22 : 5 \\
$\psi^{\prime} \stackrel{\mathrm{E1}}{\longrightarrow} 
\chi_{c1}\gamma$ & 171 & 41~(31)\footnotemark[2] & 24 : 5 \\
$\psi^{\prime} \stackrel{\mathrm{E1}}{\longrightarrow} 
\chi_{c0}\gamma$ & 261 & 46~(38)\footnotemark[2] & 26 : 5 \\
$\psi^{\prime} \stackrel{\mathrm{M1}}{\longrightarrow} 
\eta_{c}^{\prime}\gamma$ & 32 & 0.04 &  \\
$\psi^{\prime} \stackrel{\mathrm{M1}}{\longrightarrow} 
\eta_{c}\gamma$ & 638 & 0.91 & 0.75 : 0.25 \\[3pt]
$\psi(3770) \stackrel{\mathrm{E1}}{\longrightarrow} 
\chi_{c2}\gamma$ & 208 & 3.7 & \\
$\psi(3770) \stackrel{\mathrm{E1}}{\longrightarrow} 
\chi_{c1}\gamma$ & 250 & 94 & \\
$\psi(3770) \stackrel{\mathrm{E1}}{\longrightarrow} 
\chi_{c0}\gamma$ & 338 & 287 & \\[3pt]
$\eta_{c2} \stackrel{\mathrm{E1}}{\longrightarrow} \psi(3770)
\gamma$ & 45 & 0.34 & \\
%%%%%%%%%%%%%%%%%%%%%%%%%%%%%%%%%%%%%%%%%%%%%%%%%%%%%%%%%%%%%%%%%%%%%%%%%
%                                                                       %
%   $\eta_{c2} \stackrel{\mathrm{M3}}{\longrightarrow} \psi^{\prime}    %
%   \gamma$ & 127 & $10^{-4}$ & \\                                      %
%                                                                       %
%%%%%%%%%%%%%%%%%%%%%%%%%%%%%%%%%%%%%%%%%%%%%%%%%%%%%%%%%%%%%%%%%%%%%%%%%
$\eta_{c2} \stackrel{\mathrm{E1}}{\longrightarrow} h_{c}
\gamma$ & 278 & 303 & \\[3pt]
%%%%%%%%%%%%%%%%%%%%%%%%%%%%%%%%%%%%%%%%%%%%%%%%%%%%%%%%%%%%%%%%
%                                                              %
%   $\eta_{c2} \stackrel{\mathrm{M3}}{\longrightarrow} \jpsi   %
%   \gamma$ & 650 & 0.9 & \\[3pt]                              %
%                                                              %
%%%%%%%%%%%%%%%%%%%%%%%%%%%%%%%%%%%%%%%%%%%%%%%%%%%%%%%%%%%%%%%%
%%%%%%%%%%%%%%%%%%%%%%%%%%%%%%%%%%%%%%%%%%%%%%%%%%%%%%%%%%%%%%%%%%%%%%%
%                                                                     %
%   $\psi_{2} \stackrel{\mathrm{M1}}{\longrightarrow}                 %
%   \eta_{c}^{\prime}\gamma$ & 157 & \fbox{$2 \times 10^{-4}$} & \\   %
%                                                                     %
%%%%%%%%%%%%%%%%%%%%%%%%%%%%%%%%%%%%%%%%%%%%%%%%%%%%%%%%%%%%%%%%%%%%%%%
$\psi_{2} \stackrel{\mathrm{E1}}{\longrightarrow} \chi_{c2}\gamma$ & 
250 & 56 & \\
$\psi_{2} \stackrel{\mathrm{E1}}{\longrightarrow} \chi_{c1}\gamma$ & 
292 & 260 & \\
%%%%%%%%%%%%%%%%%%%%%%%%%%%%%%%%%%%%%%%%%%%%%%%%%%%%%%%%%%%%%%%%%%%%%%%%%%%%
%                                                                          %
%   $\psi_{2} \stackrel{\mathrm{M3}}{\longrightarrow} \eta_{c}\gamma$ &    %
%   744 & 0.7 & \\                                                         %
%                                                                          %
%%%%%%%%%%%%%%%%%%%%%%%%%%%%%%%%%%%%%%%%%%%%%%%%%%%%%%%%%%%%%%%%%%%%%%%%%%%%
\end{tabular}
\end{ruledtabular}
\footnotetext[1]{Derived from Ref.~\cite{Groom:2000in}.}
\footnotetext[2]{Corrected for coupling to decay channels as in 
Ref.~\cite{Eichten:1980ms}.}
\end{table}

Integrating all the calculated rates, we note that the radiative 
decays should be prominent, with branching fractions $B(h_{c} 
\rightarrow \eta_{c}\gamma) \approx \cfrac{2}{5}$, $B(\eta_{c2} 
\rightarrow h_{c}\gamma) \approx \cfrac{2}{3}$, and $B(\psi_{2} 
\rightarrow \chi_{c1,2}\gamma) \approx \cfrac{4}{5}$, of which 
$B(\psi_{2} \rightarrow \chi_{c1}\gamma) \approx \cfrac{2}{3}$.

\begin{table}%[H] add [H] placement to break table across pages
\caption{\label{table:Bdecayrates} Measured and estimated branching 
fractions for $B$ decays to quarkonium levels.
}
\begin{ruledtabular}
\begin{tabular}{ccd}
\multicolumn{2}{c}{$c\bar{c}$ state} &  
\multicolumn{1}{c}{$\Gamma(B \rightarrow (c\bar{c}) + X)/\Gamma(B 
\rightarrow \hbox{all})$ (\%)} \\
\hline
1\slj{1}{1}{0} & $\eta_{c}$ & \approx 0.53\footnotemark[1] \\
1\slj{3}{1}{1} & $\jpsi$ & 0.789 \pm 0.010 \pm 
0.034\footnotemark[2]\footnotemark[3] \\
1\slj{1}{2}{1} & $h_{c}$ & 0.132 \pm 0.060\footnotemark[4] \\
1\slj{3}{2}{0} & $\chi_{c0}$ & 0.029 \pm 0.012\footnotemark[4] \\
1\slj{3}{2}{1} & $\chi_{c1}$ & 0.353 \pm 0.034 \pm 
0.024\footnotemark[2]\footnotemark[5] \\
1\slj{3}{2}{2} & $\chi_{c2}$ & 0.137 \pm 0.058 \pm 
0.012\footnotemark[2] \\
2\slj{1}{1}{0} & $\eta_{c}^{\prime}$ & \approx 0.18\footnotemark[1] \\
2\slj{3}{1}{1} & $\psi^{\prime}$ & 0.275 \pm 0.020 \pm 
0.029\footnotemark[2] \\
1\slj{1}{3}{2} & $\eta_{c2}$ & 0.23\footnotemark[6] \\
1\slj{3}{3}{1} & $\psi$ & 0.28\footnotemark[6]  \\
1\slj{3}{3}{2} & $\psi_{2}$ & 0.46\footnotemark[6]  \\
1\slj{3}{3}{3} & $\psi_{3}$ & 0.65\footnotemark[6]  \\
\end{tabular}
\end{ruledtabular}
\footnotetext[1]{Scaled from \slj{3}{1}{1} rate.}
\footnotetext[2]{Data from {\protect{\cite{Nash:2002vr}}} and 
{\protect{\cite{Brigljevic:2001jm}}}.} 
\footnotetext[3]{Known feed-down from 2S state removed.}
\footnotetext[4]{Scaled from \slj{3}{2}{1,2} rates using 
Eqn.~({\protect{\ref{eq:thebod}}}).}
\footnotetext[5]{Known feed-down from 2S and 1P states removed.}
\footnotetext[6]{Computed; see {\protect{\cite{Yuan:1997we}}}.}
\end{table}

\textit{Charmonium production in $B$ decays.} Expectations for the 
fractions of $B$-meson decays leading to charmonium production are 
presented in Table~\ref{table:Bdecayrates}.
To estimate the $B \rightarrow \slj{1}{1}{0}$ production rates, we
appeal to the suggestion~\cite{Ko:1996iv} that the ratio of
spin-singlet to spin-triplet decay rates is relatively insensitive to
poorly calculated matrix elements,
$\Gamma(B \rightarrow n\slj{3}{1}{1} + X)/\Gamma(B
  \rightarrow n\slj{1}{1}{0} + X) = 1 + 8m_c^2/m^2_b \approx 1.5$.                                                                                                                                                                                       
The inclusive production of 1P states in $B$ decays can be
expressed~\footnote{In writing Eqn.~(\ref{eq:thebod}), we have absorbed
coefficient functions ($C_{\pm}$) and $m_B$ dependence into our
$\widetilde{H}$.} in terms of color-singlet and color-octet
contributions as~\cite{Bodwin:1992qr}:
  \begin{eqnarray}                                                                          
    \Gamma(b \rightarrow h_c + X)/\Gamma(b \rightarrow 
    \ell^{-}\bar{\nu}_{\ell}+X)        & \approx & 14.7 \widetilde{H}_8    \nonumber \\               
    \Gamma(b \rightarrow \chi_{c0} + X)/\Gamma(b \rightarrow 
    \ell^{-}\bar{\nu}_{\ell}+X)  & \approx & 3.2 \widetilde{H}_8    \nonumber \\                
    \Gamma(b \rightarrow \chi_{c1} + X)/\Gamma(b \rightarrow 
    \ell^{-}\bar{\nu}_{\ell}+X)  & \approx & 12.4 \widetilde{H}_1 + 9.3 
    \widetilde{H}_8    \nonumber \\     
    \Gamma(b \rightarrow \chi_{c2} + X)/\Gamma(b \rightarrow 
    \ell^{-}\bar{\nu}_{\ell}+X)  & \approx & 15.3 \widetilde{H}_8    
    \label{eq:thebod}                           
  \end{eqnarray}                                                                            
Using the measured rates for inclusive $\chi_{c1}$ and $\chi_{c2}$
production summarized in Table~\ref{table:Bdecayrates} we extract
$\widetilde{H}_8 = (8.95 \pm 3.79) \times 10^{-5}$ and
$\widetilde{H}_1 = (2.18 \pm 0.31) \times 10^{-4}$, which determine
the inclusive branching fractions for $\chi_{c0}$ and $h_{c}$.
No measurements exist to guide our expectations for the production of
1D states in $B$ decays, so we must rely for the moment on theoretical
calculations~\cite{Yuan:1997we} that suggest production rates roughly 
comparable to those for other charmonium states.

\textit{Observing the missing narrow states.} Radiative
transitions among charmonium levels are the key to discovering the
remaining narrow states.  Approximately 90K $B \rightarrow K\eta_{c}$
events are produced in the Belle experiment's data sample.  Using the
production rates of Table~\ref{table:Bdecayrates}, and making the
plausible assumption that $B(B \rightarrow K^{(*)} + (c\bar{c})) / B(B
\rightarrow X + (c\bar{c}))$ is universal, we estimate that 70K
$K\eta_{c}$ events are directly produced, 8.1K events arise in the
cascade $B \rightarrow K\eta_{c2} \rightarrow K\gamma(280\mev) h_{c}
\rightarrow K \gamma(280\mev) \gamma(500\mev) \eta_{c}$, and 11.7K events arise from $B
\rightarrow K h_{c} \rightarrow K \gamma(500\mev) \eta_{c}$, and that the
sample that yielded the $39 \pm 11$ $\eta_{c}^{\prime}$ discovery
events was about 30K events.  Likewise, the large radiative branching
ratios of $\psi_2$ to $\chi_{c1,2}$ and of $\chi_{c1,2}$ to $\jpsi$
provide another striking double-gamma transition with $B(B \rightarrow
X\psi_2) \sum_{J=1,2} B(\psi_2 \rightarrow \chi_{cJ}\gamma) B(\chi_{cJ}
\rightarrow \jpsi\gamma)/B(B \rightarrow X \jpsi) \simeq 0.12$.
The signal $B \rightarrow K\eta_{c2} \rightarrow K \gamma(280\mev) 
h_{c} \rightarrow K \gamma(280\mev)+\hbox{hadrons}$ may also provide a 
simultaneous observation of $\eta_{c2}$ and $h_{c}$.

We close with a few examples of the insights to be expected from the
discovery and investigation of the missing charmonium levels.  The
displacement of the 1\slj{1}{2}{1} $h_{c}$ from the 1\slj{3}{2}{J}
centroid is sensitive to Lorentz structure of the interquark
potential.  The $\psi^{\prime}$-$\eta_{c}^{\prime}$ splitting is
sensitive to a number of influences beyond simple potential models,
including the effect of virtual decay channels~\cite{Martin:1982nw}. 
The positions of $\eta_{c2}$ and $\psi_{2}$ will further constrain
analytic calculations of spin-dependent forces.  When compared with
$\psi(3770)$, they will provide another test of the influence of decay
channels in the charm threshold region.  Observation of $\psi(3770)$
in $B \rightarrow K^{(*)} D \bar{D}$ will help to calibrate expectations for
the production of the narrow states.  The same final state might yield 
evidence for $\psi_{3}$ \slj{3}{3}{3}. The details of $\psi(3770)$
decays are sensitive to S-D mixing~\cite{Kwong:1988ae}.  

\textit{Outlook:} The discovery of $\eta_{c}^{\prime}$ as a product of
$B$ decays realizes a long-held hope and raises new possibilities for 
filling out the charmonium spectrum. The \textsf{CP} violation experiments
will enrich our knowledge of $B \rightarrow (c\bar{c}) + X$, aiding
our ability to estimate the production of unknown states.

\begin{acknowledgments}
We thank Steve Olsen for posing the question that stimulated this
work.  KL's research was supported in part by a
Fermilab Frontier Fellowship and by the Department of Energy
under Grant~No.~DE--FG02--91ER40676.  Fermilab is operated by
Universities Research Association Inc.\ under Contract No.\
DE-AC02-76CH03000 with the U.S.\ Department of Energy.


\end{acknowledgments}

\bibliography{Missing}

\end{document}